\documentclass[%
    reprint,
    aps,
    prb,
    superscriptaddress,
    floatfix,    
    amssymb,
    amsfonts,
]{revtex4-2} 

\usepackage[T1]{fontenc}
\usepackage[utf8]{inputenc} 
\usepackage[english]{babel}

\usepackage{amsmath,amsthm,mathtools,mathrsfs,physics,siunitx}
\usepackage{hyperref}
\usepackage{graphicx}
\usepackage{xcolor}

\usepackage{microtype}  
\usepackage[capitalize]{cleveref}   

\newcommand{\eg}{\ensuremath{\mathrm{e}_{\mathrm{g}}}}
\newcommand{\tg}{\ensuremath{\mathrm{t}_{2\mathrm{g}}}}
\newcommand{\egrec}{\ensuremath{\tilde{\mathrm{e}}_{\mathrm{g}}}}
\newcommand{\dxy}{\ensuremath{\mathrm{d}_{\mathrm{xy}}}}
\newcommand{\dxz}{\ensuremath{\mathrm{d}_{\mathrm{xz}}}}
\newcommand{\dyz}{\ensuremath{\mathrm{d}_{\mathrm{yz}}}}

\begin{document}

\title{Correlated electronic states at a ferromagnetic oxide interface}
\author{D.\ Jones}
\affiliation{Theoretical Physics III, Center for Electronic Correlations and Magnetism, Institute of Physics, University of Augsburg, 86135 Augsburg, Germany}
\affiliation{Augsburg Center for Innovative Technologies, University of Augsburg, 86135 Augsburg, Germany}

\author{A.\ Weh}
\affiliation{Theoretical Physics III, Center for Electronic Correlations and Magnetism, Institute of Physics, University of Augsburg, 86135 Augsburg, Germany}
\author{A.\ \"Ostlin}
\affiliation{Theoretical Physics III, Center for Electronic Correlations and Magnetism, Institute of Physics, University of Augsburg, 86135 Augsburg, Germany}
\affiliation{Augsburg Center for Innovative Technologies, University of Augsburg, 86135 Augsburg, Germany}
\author{D.\ Braak}
\affiliation{Center for Electronic Correlations and Magnetism, EP VI, Institute of Physics, University of Augsburg, 86135 Augsburg, Germany}
\author{T.\ Kopp}
\affiliation{Center for Electronic Correlations and Magnetism, EP VI, Institute of Physics, University of Augsburg, 86135 Augsburg, Germany}
\author{P.\ Seiler}
\affiliation{Center for Electronic Correlations and Magnetism, EP VI, Institute of Physics, University of Augsburg, 86135 Augsburg, Germany}
\author{U. Eckern}
\affiliation{Theoretische Physik II, Institute of Physics, University of Augsburg, 86135 Augsburg, Germany}
\author{L.\ Chioncel}
\affiliation{Theoretical Physics III, Center for Electronic Correlations and Magnetism, Institute of Physics, University of Augsburg, 86135 Augsburg, Germany}
\affiliation{Augsburg Center for Innovative Technologies, University of Augsburg, 86135 Augsburg, Germany}

\date{\today}

\begin{abstract}
We propose a minimal tight-binding model for the electronic interface layer of the
LaAlO$_3$/SrTiO$_3$ heterostructure with oxygen vacancies. In this model, the effective carriers 
are subject to oxygen vacancy induced magnetic impurities.
Both the effects of random on-site potentials and Zeeman-like exchange interactions between correlated carriers and magnetic impurities are taken into account. By applying the combined coherent potential approximation (CPA) and dynamical mean-field theory (DMFT) for a ferromagnetic state, we uncover 
a disordered Fermi-liquid regime for the majority-spins and    
a low energy scale which controls the transport of the minority-spin carriers, both induced by the magnetic impurities.
\end{abstract}

\maketitle

\section{Introduction}\label{sec:intro}
The emergence of magnetism at interfaces between SrTiO$_3$ (STO) and LaAlO$_3$ (LAO) thin films is remarkable~\cite{bri07}:
this magnetism, confined to the interface, appears to be stable up to room temperature and even coexists with a superconducting state at low temperature~\cite{li.ri.11,be.ka.11}. 
In particular, it is found that 
\si{\um}-sized ferromagnetic domains 
align in a magnetic field, causing superparamagnetic behavior~\cite{li.ri.11,be.ka.11}. 
There is evidence that oxygen vacancies play a crucial role in the magnetic moment formation at these oxide interfaces~\cite{pa.ko.12a,pa.ko.12b,park.13,sall.13}. 
Oxygen vacancies serve as electron dopants, binding two electrons each, and can render 
an otherwise band-insulating environment metallic. 
A puzzling interplay between itinerant and localized electrons in STO-based 
surfaces and interfaces is indeed suggested from scanning-tunneling 
spectroscopy~\cite{brei.10,capua.12}, magnetoresistance and anomalous Hall-effect measurements~\cite{aria.11,seiler.18},  
angle-resolved photoelectron spectroscopy, as well as photoemission experiments~\cite{be.si.13,be.mu.13,st.ca.18,st.ch.19}.

Theoretical  accounts  of  realistic  LAO/STO  heterostructures are 
challenging due to the unique combination of the
interacting quantum perspective and the complex structural bulk-to-interface 
setting. However, first-principle calculations have shown that the \textit{stoichiometric} 
heterostructure is metallic and non-magnetic, with partially filled Ti $\tg$-dominated bands 
at the Fermi level~\cite{pa.ko.11,pa.ko.12b,le.bo.14}.
Calculations based on density functional theory (DFT) using 
hybrid functionals~\cite{cossu.13} confirm the results from previous DFT evaluations, but
allow for a better  control of the band gap in dependence on the thickness of the LAO overlayer.
Moreover, it has been suggested early on
from the analysis of tunneling spectroscopy data
that the electron system at the $n$-type interface is to be considered as an electron liquid due to
electronic interactions~\cite{brei.10}.

In the presence of oxygen vacancies, a multi-orbital reconstruction is identified in DFT supercell evaluations with vacancy concentrations as low as 1/8: the Ti $\eg$-states neighboring an oxygen vacancy become occupied and, with the remaining O p-orbitals, they form reconstructed $\egrec$ molecular orbitals. The two local electrons are found to be in a triplet state which signifies the possibility of local moment formation. A mechanism similar to the double-exchange induced ferromagnetism may explain the magnetic state at the interface.~\cite{pa.ko.12b}

Static correlation effects can be taken  into account by an effective Hubbard Hamiltonian treated in Hartree--Fock approximation~\cite{pa.ko.13} and
reveal some of the relevant 
aspects of interface magnetism, 
including a defect- and carrier-concentration dependent magnetic phase diagram. 
Beyond a minimal vacancy concentration of approximately $\num{0.1}\%$ robust magnetic states were identified. This finding is confirmed in a single band CPA-evaluation~\cite{pa.ko.13b} which allows to assess magnetism for low impurity concentrations not accessible in DFT calculations.

However, in the dense-defect limit, the 
physics is closer to a minimal two-orbital Hubbard model. 
Recent calculations using a combination of DFT and dynamical mean-field theory (DMFT)~\cite{si.je.17,be.le.15,le.bo.14} have shown that
in this limit, emerging ferromagnetic order in the interface TiO$_2$ layer can indeed be 
explained by effective Zener double-exchange between an $\egrec$ orbital and an in-plane $\tg$ $\dxy$-orbital.

In this work, we propose a minimal tight-binding model for the physics of oxygen-vacancy induced magnetism at the LAO/STO interface. As an alternative to the existing supercell DMFT computations~\cite{be.le.15}, we use a combination of the coherent potential approximation (CPA)~\cite{ve.ki.68}, treating disorder through averaging, with DMFT, thereby including correlation effects~\cite{MV89a,ko.vo.04}. 
It is known that the CPA provides the exact solution for non-interacting fermions with diagonal (local) disorder on any lattice in the limit $Z \to \infty$, where $Z$ is the coordination number, provided the appropriate quantum scaling of the hopping amplitude is employed~\cite{vlaming92}. 
Its generalization to \emph{interacting} disordered electrons, i.e.\ a combination of CPA and DMFT,
has been achieved before~\cite{Janis91,ja.vo.92,ge.ko.92,Jarrell.92,pr.ja.95,ul.ja.95,ge.ko.96}. 
In our work, we pursue such a combined CPA+DMFT scheme and investigate the effects of the correlated fermions in the Zeeman field induced by the magnetic impurities at the LAO/STO interface.

Electronic correlations in fully polarized magnetic systems have been studied extensively in the context of spintronics~\cite{ka.ir.08} and it is well known that correlation induced spin fluctuations play a crucial role in metallic ferromagnets~\cite{moriya2012}.
In particular, the scattering of charge carriers on such magnetic excitations is expected to control the macroscopic properties of these materials, including transport.
Contrary to the itinerant ferromagnets, in which states near the Fermi level have quasiparticle character for both spin projections, incoherent non-quasiparticle (NQP) states appear in half-metallic alloys.
These NQP states occur near the Fermi level in the energy gap~\cite{Edwards_1973,ir.ka.83,ir.ka.94}. The density of the these states vanishes at the Fermi level, but is strongly enhanced at an energy scale of the order of the  characteristic magnon-frequency. 
This has been calculated from first principles for a prototype half-metal ferromagnet, NiMnSb~\cite{ch.ka.03}, as well as for other Heusler alloys~\cite{ch.ar.06}, zinc-blende structure compounds~\cite{ch.ka.05,ch.ma.06} and CrO$_2$~\cite{ch.al.07}. 
The origin of these states is related to ``spin-polaron'' processes~\cite{ir.ka.83,ir.ka.94} where minority spins are coupled to collective (magnon) excitations. Just recently, polaronic effects have been discussed for LAO/STO interfaces~\cite{st.ch.19}. 
It is the purpose of the present paper to analyze possible departure from the Fermi-liquid theory that may form at the LAO/STO interface in the presence of magnetic impurities.

\section{Interface Model Hamiltonian}
\label{sec:model_cpa+dmft}
\begin{figure}[ht]
    \includegraphics[width=1.0\linewidth,clip=true]{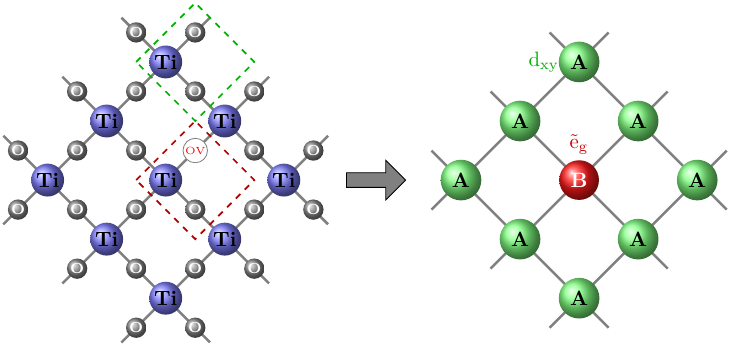}
    \caption{
        Left: The interface TiO$_2$-layer containing randomly distributed oxygen vacancies (OV). Right: Mapping onto an effective model of the molecular orbitals type-$\rm A=O-Ti-O$ (green) and type-$\rm B= O-Ti-VO$ (red). A ``local moment'' is induced by vacancies at the B-sites.
    }
    \label{model}
\end{figure}

To visualize our partitioning of the interface layer, we sketch in Fig.~\ref{model}, on the left-hand side, the interface plane that contains an oxygen vacancy (OV) situated between two adjacent Ti atoms. 
On the right-hand side we partition the lattice into $\rm O-Ti-O$ units, the green area, and the $\rm O-Ti-OV$ unit contained in the red area. 
For a larger vacancy concentration, the number of red areas/units is larger. 
This partitioning naturally leads to consider the green units ($\rm O-Ti-O$) forming the ``host'' while the vacancy containing unit as ``guest'' of the interface plane. 
By introducing the notation A for the host unit ($\rm O-Ti-O$) and B for the vacancy containing unit ($\rm O-Ti-OV$), we can map the interface plane into a substitutional alloy model of type A$_{1-c}$B$_c$.

The electronic states of the alloy component A are described by the ``molecular'' orbitals of a TiO$_2$ unit, while in the presence of vacancies B alloy component states are of TiO-molecular orbital type.  
For both molecular orbital states  (TiO$_2$/TiO)  
the symmetry breaking of the interface lowers the energy of the Ti  $\dxy$ orbital relative to
the $\dxz$ and $\dyz$ states.
The Ti $\dxz$ and $\dyz$ orbitals remain unoccupied, and a single orbital of  $\dxy$-character at each Ti-site is partially filled.
In the presence of vacancies, the TiO-molecular orbitals will carry randomly distributed magnetic impurities resulting from the $\eg$ states hybridizing with the remaining oxygen p-orbitals ($\egrec$ orbitals)~\cite{pa.ko.13,le.bo.14}.

The majority of material specific modeling approaches of the electronic liquid at the LAO/STO interface takes the DFT results as a starting point.
Thus, a low-energy effective two-orbital ($\dxy, \, \egrec$) model is projected out of (downfolded from) the DFT manifold of states.
At a second step the Hubbard interaction terms, parameterized by $U, \, U^\prime$ and $J$ are added to the downfolded hopping matrix elements and the resulting two-band Hubbard model is solved by various many-body techniques such as DMFT. The validity of such a Hubbard formulation for a spin-polarized ground state of the LAO/STO interface is hardly 
discussed in the literature; perhaps 
most notable is to trace it to a double exchange Zener-like mechanism between magnetic domains seen as impurities resulting from the oxygen vacancies.~\cite{si.je.17,be.le.15,le.bo.14} 
Alternatively to this, note that it is in fact possible to stabilize a ferromagnetic solution within the Hubbard model under the conditions: (i) a density of states with large spectral weight near the band edges~\cite{ulmk.98}, and (ii) Hund's rule coupling for the degenerate case~\cite{vo.bl.99}.

In our work we focus on the low density distribution of magnetic impurities for which a stable ferromagnetic ground state is produced by a double-exchange Zener-like mechanism. 
The starting point in our numerical analysis is the Hamiltonian written as:
\begin{equation}
    \label{eq:AMHamiltonian}
    \begin{aligned}
      \hat{\mathcal{H}} &=
      \sum_{i,\sigma} v_{i,\sigma} \hat{n}_{i,\sigma} 
      -\sum_{\langle ij \rangle,\sigma} \!\!  t_{ij}\, \hat{c}^{\dagger}_{i,\sigma} \hat{c}_{j,\sigma}^{}
      +  \sum\limits_i U_i \hat{n}_{i,\uparrow} \hat{n}_{i,\downarrow}
      \\
      v_{i,\sigma} &= \varepsilon_i  - \delta_{i,{\rm B}} \,\mathcal{B}\, \sum_{\alpha,\beta}  \hat{c}^{\dagger}_{i,\alpha} \sigma^{\alpha\beta}_z  \hat{c}_{i,\beta}^{}  -\mu \\
      U_i  &=  U_{\rm A} \delta_{i,{\rm A}} + U_{\rm B} \delta_{i,{\rm B}}
    \end{aligned}
\end{equation}
where $ \hat{n}_{i,\sigma}=\hat{c}^{\dagger}_{i,\sigma} \hat{c}_{i,\sigma}^{}$ 
are the local number operators for electrons with spin $\sigma$ at site $i$,  and $t_{ij}$ is the nearest-neighbor hopping
parameter. The on-site Coulomb interaction $U_i$ between the carriers at the A/B-molecular orbital is $U_{\rm A/B}$ and treated at the DMFT level. 
The chemical potential is $\mu$ and $\varepsilon_{i}=\varepsilon_{\rm A/B}$ 
represent the local on-site energies of the type A ($\rm O-Ti-O$) and B ($\rm O-Ti-VO$)
molecular orbitals in the binary-alloy picture.
Within our simplified model, all hopping amplitudes are taken to be site- and orbital-independent:
$t_{ij}\equiv t$.
The second term in $v_{i,\sigma}$, see Eq.~(\ref{eq:AMHamiltonian}), reflects the Hund's coupling between the electrons and the localized spins at the B-sites.
We treat these spins at the mean-field level, replacing them by a constant Zeeman field $\mathcal{B}$
at the (random) site $i$.
In principle, $\mathcal{B}$ may be obtained self-consistently as a function of the $\egrec$ orbital occupation.

Note that the index $z$ at the Pauli matrix in Eq.~(\ref{eq:AMHamiltonian}) 
refers to the direction of the local moments
which generate this Zeeman field and are usually oriented in-plane.~\cite{li.ri.11,be.ka.11}
In this realization of the interface physics, a Kondo singlet formation is excluded because we have spin-one impurities (the impurity would be only partially screened).
Finally, disorder averaging is performed through a CPA+DMFT bath.

%

%

\section{CPA+DMFT computations for the Hubbard model}

In our approach, we simplify the model by representing the effective impurity spins 
as a homogeneous external magnetic field acting only at the impurity sites.
Furthermore, we consider a square lattice with the energy scale set to $4d t^2 = {t^*}^2 $ required for the non-trivial result as $d\rightarrow \infty$
and with the half-bandwidth $D=2t^*=\SI{2}{\electronvolt}$. 
%
Within the binary alloy picture the A-type orbitals with the on-site energy $\varepsilon_{\rm A}=0.6 D$ represents the host while the amount of $c=20\%$ magnetic impurities 
\footnote{This amount of impurities is related to an approximate concentration of oxygen vacancies which itself depends on the sample preparation. Here we assume a low concentration of oxygen vacancies in order to keep a low filling consistent with most experiments.} 
are modeled by the B-type orbitals having the on-site energy $\varepsilon_{\rm B}=-0.6 D$ and being split by an exchange parameter $\mathcal{B}=0.4 D$.
\footnote{This value of $\mathcal{B}$ is estimated from the DFT results:~\cite{pa.ko.12b} the magnetic splitting of the $\dxy$ bands is approximately $\SI{1.04}{\electronvolt}$ and the corresponding hopping parameter is $\SI{0.28}{\electronvolt}$.}

We start by solving the model, Eq.~(\ref{eq:AMHamiltonian}), in the absence of the interaction term ($U=0$). The  magnetic field $\mathcal{B}$  
can be included into the random potential 
$v_{i,\sigma}$ and treated within CPA\@. According to 
standard multiple scattering theory~\cite{goni.92}, the Hamiltonian can be divided into an unperturbed reference part and 
a perturbative term, $\hat{\mathcal{H}}^{\text{ref}} + \hat{\mathcal{H}}^{\text{per}}$.
The reference Hamiltonian is
\begin{equation} 
    \hat{\mathcal{H}}^{\text{ref}}(E) = \sum_{\mathbf{k},\sigma}\left[\epsilon_{\mathbf{k}}+\Sigma_{c,\sigma}(E) \right] \hat{c}^{\dagger}_{\mathbf{k},\sigma} \hat{c}_{\mathbf{k},\sigma}^{}
\end{equation}
in reciprocal space for an electron with dispersion $\epsilon_{\mathbf{k}}$, moving 
in the energy-dependent effective medium described 
by a spin-dependent complex coherent potential $\Sigma_{c,\sigma}(E)$. 
The perturbative term is written as sum over all lattice sites,
$\hat{\mathcal{H}}^{\text{per}} = \sum_{i,\sigma}  w_{i,\sigma}$ 
with 
\begin{equation}
    w_{i,\sigma} = -\delta_{i,B}\mathcal{B}\sum_{\alpha,\beta}  \hat{c}^{\dagger}_{i,\alpha} \, \sigma^{\alpha\beta}_z  \hat{c}_{i,\beta}^{}  + 
    \left[ \varepsilon_i-\mu - \Sigma_{c,\sigma}(E) \right] \hat{n}_{i,\sigma}^{}\nonumber     
\end{equation}

Next, using the reference Green's function, $G_{c}(E)$, 
the $t$-matrix due to the scattering of the impurity spins and impurity potentials embedded 
in the effective medium is $t^{(l)}_{i,\sigma} = v^{(l)}_{i,\sigma} [1-G_{c,\sigma} v^{(l)}_{i,\sigma}]^{-1}$.
The superscript $l$ refers to the alloy components A ($\dxy$) or B ($\egrec$). 
Note that $t^{(l)}_{i,\sigma}$ describes 
the complete scattering associated with the isolated potential in the effective medium and the reference
system, and thus the reference Green's function does not contain the magnetic field contribution.
The total scattering operator is then expressed in the multiple scattering series
$
T^{(l)}_{\sigma} = \sum_i t^{(l)}_{i,\sigma} + \sum_i t^{(l)}_{i,\sigma} G_{c,\sigma} \sum_{j \ne i} t^{(l)}_{j,\sigma} + \, \dots
$
and the total Green's function 
is $G^{(l)}_\sigma = G_{c,\sigma}+G_{c,\sigma} T^{(l)}_{\sigma} G_{c,\sigma}$. Within the single-site approximation, the condition of 
$\expval*{t^{(l)}_{i,\sigma}}=0$
for any site $i$ leads to $\expval*{T^{(l)}_{\sigma}}=0$ and thus $\expval*{G^{(l)}_{\sigma} } \approx G_{c,\sigma}$, representing the
CPA. The local part of the reference (spin-polarized) Green's function is:
\begin{equation}
    \label{gc}
  G_{c,\sigma}(E) \!= \! \frac{1}{N}\sum_{\mathbf{k}} \frac{1}{E-\epsilon_{\mathbf{k}} - \Sigma_{c,\sigma}} =\! \int\limits_{-D}^{D} \frac{\rho(\epsilon)}{E-\epsilon - \Sigma_{c,\sigma}(E)}  d \epsilon
\end{equation}


We consider next the electronic correlations modeled by the presence of a Hubbard parameter $U=U_{\rm A}$  acting upon the host
orbitals~$\dxy$ only. 
For most evaluations, we choose $U=t^{*}$, consistent with tunneling spectroscopy data.~\cite{brei.10}.
We start the self-consistency loop with a guess for the self-energy $\Sigma_{c,\sigma}$, which includes 
both, the disorder and correlation effects.
The local Green's function is computed from the electronic dispersion $\epsilon_{\mathbf{k}}$ (eigenstates of the lattice Hamiltonian in the absence of disorder and electronic correlations) and the initial guess for the self-energy $\Sigma_{c,\sigma}$ (Eq.~\ref{gc}). From the coherent (local) Green's 
function, alloy component ($l=\egrec$ and $\dxy$) Green's functions are computed as 
\begin{equation}\label{eq:c2}
G_{l,\sigma} = G_{c,{\sigma}} \left[1-(v_i^{(l)}+\Sigma^{{\rm DMFT} }_{l,\sigma}\!-\Sigma_{c,\sigma}) \,G_{c,\sigma} \right]^{-1}
\end{equation}
for a given (fixed) disorder realization. In the next step, the many-body problem is solved using 
the DMFT methodology: the DMFT bath Green's function is constructed as
\begin{equation}\label{eq:c3}
(\mathcal{G}_{l,\sigma})^{-1}=(G_{l,\sigma})^{-1}+\Sigma^{\rm DMFT}_{l,\sigma}.
\end{equation}
for each spin component and given the $\Sigma^{\rm DMFT}_{l,\sigma}[\mathcal{G}_{l,\sigma}]$ we request that the alloy components should fulfill the CPA equation: $G_{c,\sigma}=\sum_l c_{l}\, G_{l,\sigma}$.
We solve the impurity problem using the CT-HYB quantum Monte Carlo (QMC) impurity solver as implemented in TRIQS/cthyb~\cite{triqs.cthyb}. 
Since Monte-Carlo based impurity solvers notoriously produce noisy self-energies computed via a direct Dyson equation, we employ the
constrained residual minimization (CRM) method~\cite{la.ka.25} to stabilize the impurity self-energy $\Sigma^{\rm DMFT}_{l,\sigma}$.
%
The coherent and spin-resolved Green's function $G_{c,\sigma}$ corresponds to a disorder averaging over the disorder realizations with concentrations $c_{\dxy}=80\%$ and $c_{\egrec}=20\%$. 
From the newly computed $G_{c,\sigma}$, the coherent self-energy $\Sigma_{c,\sigma}$ follows directly. To close the self-consistency loop, $G_{c,\sigma}$ and $\Sigma_{c,\sigma}$ are returned into Eq.~(\ref{eq:c2}) to produce new alloy component Green's functions.
On a more formal level, this algorithm has been presented in Refs.~\onlinecite{ul.ja.95,os.vi.18}.

\section{Results}

\subsection{Spectral functions and Self-energies}

\begin{figure}[ht]
    \includegraphics[width=1\linewidth,clip=true]{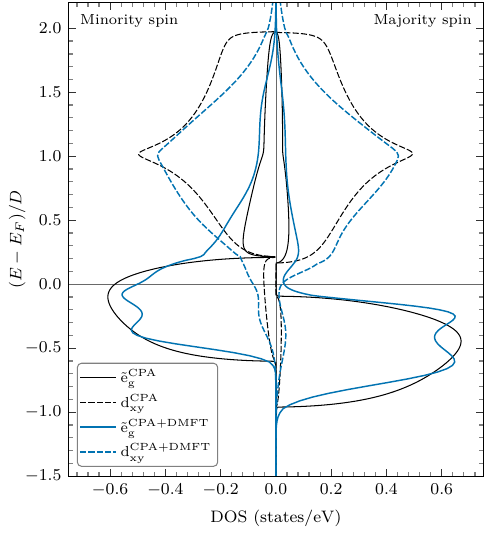}
    \caption{
        Alloy component DOS ($\dxy$-$80\%$-dashed lines and $\egrec$-$20\%$-solid lines). Non-interacting results $U=0$ (black) and $U=0.5D = t^*$ (blue) at the inverse temperature $\beta=20 / D$.}
    \label{DOS_CPA_DMFT}
\end{figure}

In Fig.~\ref{DOS_CPA_DMFT} (black lines) we present numerical results
for the spin-polarized density of states (DOS) in the dilute case $c=20\%$ of $\egrec$-orbitals for the model Hamiltonian Eq.~(\ref{eq:AMHamiltonian}) in the absence of interactions ($U=0$).
Since the CPA does not involve many-body calculation the spectra can be obtained directly on the real-frequency (energy) axis.
The CPA solution produces a fully spin-polarized ground state, sometimes called a majority-spin half-metallic ferromagnetic ground state~\cite{ka.ir.08,ch.ar.09}.
The majority-spin $\egrec$-orbital spectral function forms a split-off band, effectively separated from the main $\dxy$-host band. 
The Fermi energy is situated in the minority spin channel, in the vicinity of the maxima
of the split-off impurity band DOS thus, CPA yields minority-spin metallic, magnetic bands.
Note that the CPA effective medium has a finite energy support, limited by the common bandwidth~\cite{we.zh.21}, according to the Gershgorin theorem~\cite{gers.31}.
%

Including electronic correlations on the host $\dxy$ orbitals causes a considerable broadening of the host electron's DOS (blue solid line) in both spin channels. The minority-spin spectral function of the host orbital gains weight at $E_F$.
Both majority and minority-spin $\egrec$-orbitals spectral functions are slightly shifted towards the Fermi level.
In the vicinity of the $\dxy$ majority-spin band-edge, a shoulder is formed and its tail cross the Fermi energy and contributing to depolarization. 
Since the DMFT impurity problem is solved using a QMC solver, which yields Green’s functions on the Matsubara axis, we performed the analytical continuation from Matsubara frequencies to the real-frequency axis using the Maximum Entropy (MaxEnt) method as implemented in TRIQS/maxent.~\cite{triqs.maxent}.
Important to notice that although only the host $\dxy$-orbitals are subject to electronic correlations, the non-correlated $\egrec$ spectral functions are significantly modified through the combined CPA+DMFT self-consistency.


In the following we analyze the behavior of the self-energy, which represents a self-consistently determined homogeneous complex and energy dependent effective medium that carries information about both disorder and correlation.
Analytical continuation of the self-energy was performed using the same MaxEnt algorithm used for the spectral functions.

\begin{figure}[ht]
    \includegraphics[width=\linewidth,clip=true]{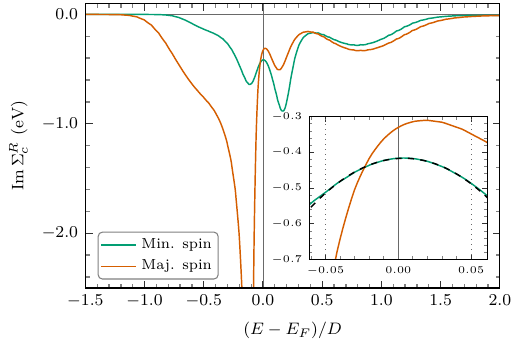}
    \caption{
        The imaginary part of the retarded self-energy $\Sigma^R_{c}$ computed using CPA+DMFT for $U=0.5D = t^*$ and $\beta=20/D$. The inset shows the self-energy around the Fermi level. 
    } 
    \label{Sigma}
\end{figure}

The main plot of Fig.~\ref{Sigma} shows the imaginary part of the CPA+DMFT coherent self-energy. 
The inset shows the self-energies in a limit range $E_F \pm k_BT$.  
Near the Fermi level, we find that the minority-spin channel coherent self-energy exhibits an imaginary part that can be excellently described using a quadratic energy dependence, i.e., $\Im \Sigma^R_{c} \propto (E-E_F)^2$ (dashed curve in the inset of Fig.~\ref{Sigma}).
This enforces the Fermi liquid description for the minority spin electrons. 
Although the system remains a Fermi liquid, the imaginary part of the self-energy does not vanish at $E_F$; its finite value reflects the scattering rate in the combined disordered and correlated medium. Ultimately, this leads to an energy-dependent broadening of the minority-spin electronic bands.
%
The departure from conventional Fermi-liquid behavior becomes evident in the majority-spin channel. The inset shows a shift relative to the Fermi energy $E_F$. Nevertheless, the majority-spin self-energy  can still be fitted by an approximately parabolic dependence.

\begin{figure}[ht]
    \includegraphics[width=\linewidth,clip=true]{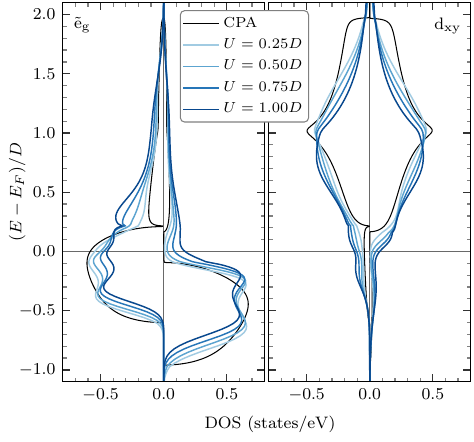} 
    \caption{
        Orbital resolved density of states for various values of $U$ computed with CPA+DMFT at the inverse temperature $\beta=20 / D$.
    }
    \label{dos_U_sigma}
\end{figure}

Figure~\ref{dos_U_sigma} represents the spin and orbital resolved spectral functions for various $U$ values starting from the uncorrelated CPA up to the most strongly correlated case, when the strength of local interaction equals the half bandwidth ($U = D$). We observe a continuous filling up of the majority-spin CPA gap with increasing $U$. 
Most notably, the position of the majority-spin interacting band-edge states approaches the Fermi level leading to a more important spectral weight at $E_F$ and stronger depolarization effects. 

\subsection{Impurity state conductivity computation}

The behavior of the single-particle density of states and self-energies was discussed in the previous section.
In particular, the emergence of states in the majority-spin channel of host $\dxy$-orbitals is a clear signature of important electronic correlations leading to the closure of the CPA-gap. 
Here we address how the combined effects of disorder and electronic correlations manifest in charge transport, focusing explicitly on the resistivity. 
Our analysis focuses on the impurity-induced $\egrec$ orbitals, since the host electrons remain Fermi-liquid quasiparticles even in the presence of disorder and correlations. 
The resistivity therefore reveals changes in the quasiparticle properties of these impurity states.
For this purpose, we compute in the following the conductivity in the limit of infinite coordination~\cite{ge.ko.96}. 
%
In this simplified setup, the conductivity for the $\egrec$-band is given by the convolution of the spectral functions~\cite{pr.ja.95,ge.ko.96,pr.co.93,ja.fr.95}:
\begin{equation} \nonumber 
    \sigma(\omega) / \sigma_0 = \int_{-\infty}^{\infty} dE \frac{f(E)-f(E+\omega)}{\omega} A_{\egrec}(E) A_{\egrec}(E+\omega),
\end{equation}
where $A_{\egrec}(E)$ is the interacting spectral function of the impurity orbital $\egrec$.
\begin{figure}[ht]
  \includegraphics[width=\linewidth,clip=true]{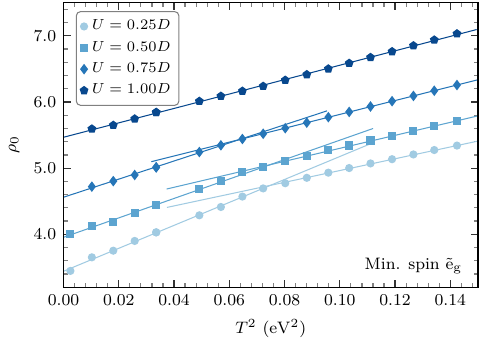}
  \caption{
    The inverse of the Drude peak $\rho_0 = 1 / \sigma(0, T)$ in dependence on temperature $T^2$ for the minority spin channel of the guest $\mathrm{\tilde{e}_{g}}$ orbital (B) computed with CPA+DMFT. 
    The solid lines are linear fits of $\rho_0$. 
    }
    \label{rho_T}
\end{figure} 

In Fig.~\ref{rho_T} we show the computed DC resistivity $\rho_0(T)$ as function of temperature for a range of $U$ values. 
We now relate these results to the behavior of the $\egrec$-orbital spectral function.
The spectral function at the Fermi level $A_{\egrec}(E_F)$, shown in Fig.~\ref{dos_U_sigma}, decreases with increasing $U$.
As long as the system remains metallic, this reduction corresponds to a decreasing quasiparticle weight $Z$, while the spectral function simultaneously develops Hubbard bands.
In correlated metals, the DC conductivity in general scales as $\sigma_{0} \propto Z \tau$~\cite{pr.ja.95,ge.ko.96,pr.co.93,ja.fr.95}. 
In addition to correlations, disorder further amplifies scattering processes, i.e., correlation-induced scattering on top of impurity scattering.
Consequently, the quasiparticle lifetime $\tau$ is reduced, leading to a decreasing $\sigma_{0}$ and hence and increasing DC resistivity, as observed in Fig.~\ref{rho_T}.
Strictly speaking, at $T=0$ the DMFT imaginary part of self-energy vanishes, implying a zero scattering rate. Nevertheless, the disorder induced CPA self-energy, which contributes to the total effective medium self-energy (see Fig.~\ref{Sigma}), is expected to remain finite.

In the following we comment the temperature dependence of the resistivity $\rho_{0} =1/\sigma(0)$.
The relation $\rho_{0}(T) \propto T^2$ is specific of a Fermi liquid at low temperatures.  
For the minority $\egrec$-orbital states, we clearly identify a kink in the linear behavior of $\rho_{0}(T)$ as a function of $T^2$ for values of the interaction parameter smaller than the half bandwidth: $U < D$. At $U=D=2t^*$ the kink disappears and $\rho_{0}$ follows a straight line down to the lowest (numerically achievable) computed temperature. 
The change in the slope implies the existence of an energy scale that affects the low temperature transport of the minority carriers. 
This energy scale, $k_{\rm B} T^\star$, is controlled by the frequency dependent self-energy
at the Fermi energy (i.e., the scattering amplitudes), 
the bandwidth, and the magnitude of the exchange parameter $\abs{\mathcal{B}}$.
A more in depth analysis is required to reveal the nature and the robustness of this energy scale, which is beyond the scope of the present analysis. 


\section{Discussion and Conclusions}
Ferromagnetic small-sized domains at the LAO/STO interface have been established experimentally~\cite{li.ri.11,be.ka.11} and traced back to the existence of oxygen vacancies.
Thus, a minimal model has to be able to capture both electronic correlations and randomness in configurations of vacancies  present at the interface.  
Such a two-orbital ($\egrec$--$\tg$) effective model has been previously formulated and the results were interpreted in terms of an effective double exchange between nearly localized 
$\egrec$ and mobile $\dxy$ electrons. Using a DFT+DMFT~\cite{le.bo.14} technique for supercell structures and very precise QMC~\cite{le.bo.13,le.bo.14} computations (as well as slave boson techniques~\cite{be.le.15}), a strongly re-normalized temperature-dependent quasiparticle weight as well as a significant incoherent spectral
weight were obtained. It was previously also emphasized that oxygen vacancies are responsible for the completely different behavior of weakly correlated $\tg$ low-energy quasiparticles and localized ``in-gap'' states of dominant $\eg$ character~\cite{pa.ko.12b,le.je.16}. This dichotomy in the behavior of various orbitals at the interface has been supported through experimental studies~\cite{al.je.16,st.ch.19}.

Based on these previous results and observations, the focus of our study is to characterize the nature of the spin-polarized electronic states in such a minimal model, including the averaged effect of disorder through the coherent potential approximation on the same footing with electronic correlations.  
The supercell computations reported previously correspond to a specific disorder realization (i.e., a particular vacancy configuration) and therefore do not capture disorder-averaging effects. 
As temperature is lowered, the Hubbard model of the two effective orbitals~\cite{le.bo.14} is found to be a Fermi liquid, which is to be expected in the infinite-$d$ limit. 
Complementary to this, our results show that the minority-spin electronic states retain Fermi-liquid properties, whereas in the majority-spin channel we find a disordered Fermi-liquid regime.
In this regime, quasiparticles are broadened but not destroyed. Further analysis is required to assess whether a singular frequency dependence or a breakdown of the quasiparticle residue occurs, which would constitute evidence for non-Fermi-liquid behavior.

In line with this, we identify oxygen vacancies as the source of the local classical spin states, which we model in the ferromagnetic regime by an effective magnetic field $\mathcal{B}$.
We note that even small external fields of $\SIrange{0.01}{0.02}{\tesla}$ are sufficient to align the magnetic domains in the investigated samples.~\cite{li.ri.11}
Thus, the ferromagnetic regime studied here is experimentally accessible, even if the system is globally superparamagnetic. Moreover, the ferromagnetic state remains robust above a minimal vacancy concentration of $\num{0.1}\%$.~\cite{pa.ko.13}.

The disorder problem solved within CPA yields a fully spin-polarized electronic state.
In spin-polarized systems, a signature of electronic correlations is the emergence of non-quasiparticle states~\cite{ka.ir.08} in the spin channel that is nearly empty or nearly fully occupied.
In our modeling of the LAO/STO interface, the majority-spin channel provides precisely these conditions; accordingly, we characterize the resulting electronic states as belonging to a disordered Fermi-liquid regime.
To characterize the transport properties of the minority-spin electronic states, we employed a simplified formalism to compute the optical conductivity as a function of temperature. A kink in the DC resistivity signals a renormalization of the quasiparticles, in qualitative agreement with previous DFT+DMFT calculations~\cite{le.bo.14}.
Thus, for minority-spin $\egrec$-orbital electrons we identify a crossover between two transport regimes, reflected by different slopes in the $T^2$ dependence of the DC resistivity $\rho_0(T)$. Within a Fermi-liquid description, this corresponds to distinct transport relaxation rates.
For stronger correlations, i.e., for $U$ of the order of the bandwidth, the kink disappears and a single relaxation regime emerges; its nature will be addressed in a subsequent work.
Note, however, that we observe depolarization accompanied by the emergence of majority-spin charge carriers.
Our results suggest that these anomalous properties are governed not only by scattering processes, but also by the effective magnetic field $\mathcal{B}$, which originates from Hund’s-rule coupling and sets a low-temperature energy scale.
We propose that this scale could be probed in transport measurements.

We expect that the approach presented here is also applicable to hydrogen dopants in the TiO$_2$ interface layer, where an electron is transferred from the H $1\mathrm{s}$ orbital to Ti $\dxy$ orbitals, enabling the formation of interface-confined magnetic moments as recently reported~\cite{pi.ey.19}.

%
%
%

\begin{acknowledgments} 
This work was supported by the Deutsche Forschungsgemeinschaft (DFG, German Research Foundation)–TRR 360–492547816. 
\end{acknowledgments}

\bibliography{main}

\end{document}